\documentstyle[12pt] {article}
\begin{document}

\title{Relativistic meson spectroscopy and in-medium effects}
\author{Alan J. Sommerer, A. Abd El-Hady, and John R. Spence\\
{\it Department of Physics and Astronomy, Iowa State University,} \\
{\it Ames, IA  50011}\\ \\
and James P. Vary\\
{\it Department of Physics and Astronomy, Iowa State University,}\\
{\it Ames, IA 50011 and} \\
{\it Institute for Theoretical Physics, University of Heidelberg,} \\
{\it D-69120 Heidelberg, Germany.}}

\maketitle

\begin{abstract}
We extend our earlier model of $q\bar q$ mesons using relativistic
quasipotential (QP)
wave equations to include open-flavor states and running quark-gluon
coupling effects.  Global fits to meson spectra are achieved with rms
deviations from experiment of
43-50 MeV.  We examine in-medium effects through their influence on
the confining interaction and predict the confining strength at which
the masses of certain
mesons fall below the threshold of their dominant decay channel.
\end{abstract}


\baselineskip .75cm
\pagebreak

Interest in calculating the properties of hadronic matter has grown in
recent years because it is believed that understanding the behavior of
these properties in high density and temperature environments is necessary for
interpreting the results of relativistic heavy ion collisions.  It is
expected that
observables such as masses, widths, and couplings will change in hot, dense
environments like those achieved in experiments attempting to create
conditions for the phase transition to the quark-gluon plasma [1].

Several papers have treated constituent quark models using two-body
relativistic wave equation treatments [2,3].  We here improve and
extend the model of Ref. [3] by including open-flavor mesons in the
spectrum fits and incorporating a running quark-gluon coupling.
We then provide initial results for in-medium effects through their
influence on the confining interaction.

The interaction kernel consists of a one-gluon exchange interaction in the
ladder approximation, $V_{OGE}$, and a phenomenological, long-range
confinement potential, $V_{CON}$.   We
treat the potentials in momentum-space according to the methods
described in Ref. [4]. This interaction takes the form
$$V_{OGE}+V_{CON}={4\over 3}\alpha_s{\gamma_\mu\otimes\gamma_\mu\over
{(q-q')^2}}
+\sigma{\rm\lim_{\mu\to 0}}{\partial^2\over\partial\mu^2}
{{\bf 1}\otimes{\bf 1}\over-(q-q')^2+\mu^2}.\eqno(1)$$
Here, $\alpha_s$ is the strong coupling, which is weighted by the meson color
factor of ${4\over 3}$, and the string tension $\sigma$ is the strength of
the confining part of the interaction.  We adopt a scalar Lorentz
structure $V_{CON}$ since this choice is supported by lattice
results [5] and by phenomenology, e.g. [3].

There are several choices of relativistic two-body wave equations which can
be used to treat the interactions of Eq. (1).  In this
investigation, we employ two different equations, a spinor
version of the Thompson equation [6] and the equation introduced in
Ref. [3].  For convenience, we will refer to these equations as ``reduction
A'' and ``reduction B'' respectively.
These two integral equations result from different choices of
three-dimensional propagators used to reduce the Bethe-Salpeter equation
from four to three dimensions.

Assuming a bound state of quarks
with equal masses in the center-of-mass $0^-$ channel, the reduction A
equation reads

$$(2\omega-E)\,\psi^0_+(q)={4\alpha_s\over 3\pi q}I_1(q,q')+{\sigma\over
2\pi q}I_2(q,q'),\eqno(2)$$

and the reduction B equation reads
$$(4\omega^2-E^2)\,\psi^0_+(q)={8\alpha_s E\over 3\pi
q}I_1(q,q')+{\sigma
E\over
\pi q}I_2(q,q').\eqno(3)$$
In each case
$$I_1(q,q')=\int^\infty_0
dq'q'Q_0(Z){2\omega\omega'-m^2\over \omega\omega'}\psi^0_+(q'),\eqno(4)$$
and
$$I_2(q,q')={\rm \lim_{\mu\to
0}}{\partial^2\over\partial\mu^2}\int^\infty_0
dq'q'
\biggl\{{Q_0(Z'){-\omega\omega'-m^2\over\omega\omega'}+
\biggl[Z'Q_0(Z')-1\biggr]{qq'\over\omega\omega'}\biggr\}}
\psi^0_+(q').\eqno(5)$$
The quantities $q$ and $q'$ are magnitudes of relative three-momenta,
$\omega^{(')}=[m^2+q^{(')2}]^{1/2}$, $m$ is the constituent quark mass,
$Q_0$ is the
Legendre function of the second kind with argument $Z=(q^2+q'^2)/2qq'$
or $Z'=(q^2+q'^2+\mu^2)/2qq'$, and
$E$ is the energy to be solved for along with the amplitudes $\psi^0_+$.
The `+' subscript on the amplitudes indicates that only the positive energy
components have been retained in these equations.

We have improved the calculated spectra of Ref. [3] by incorporating
a running strong coupling into the model.  Specifically, we have
constrained the coupling to
run as in the leading log expression for $\alpha_s$,
$$\alpha_s(Q^2)={4\pi\alpha_s(\mu^2)\over 4\pi+\beta_1\alpha_s(\mu^2){\rm
ln}\bigl({Q^2/\mu^2}\bigr)},\eqno(6)$$
where $\beta_1=11-2n_f/3$ and $n_f$ is the number of quark flavors.
The most precise measurements of $\alpha_s$ have been at $M_Z$, the mass
of the Z-boson [7]:
$\alpha_s(\mu^2=M_Z^2)\simeq 0.12$, and the running couplings used in
the
following
calculations are anchored to this point.
We have elected  to relate $Q^2$ to the meson mass scale through
$$Q^2 = \xi^2 M_{meson}^2 + \rho^2,\eqno(7)$$
where the parameters $\xi$ and $\rho$ are determined by the fit.  The second
term, if it is large enough, ensures a finite saturation value of
$\alpha_s$ as $Q^2$ goes to zero.

We have also expanded the work of Ref. [3] by including open-flavor mesons
in our fit.

The constituent masses, the string tension $\sigma$, and the
parameters $\xi$ and $\rho$ are adjusted to minimize the RMS
deviation between the calculated meson masses and the experimental masses
listed in the Particle Data Tables [8].

There is the well debated issue of whether the low-mass pion (and
later, when open-flavor states are considered, the $K$ meson)
should be included in a constituent quark model picture, even a relativistic
one.  We adopt the philosophy that if a model adequately reproduces
the $\pi$ and $K$
masses, then these states may be retained in the
calculations using that model.  In the present case, reduction A does
poorly at representing these low-mass mesons, and these states are therefore
not included in calculations using reduction A.

The global fits includes 45 states for reduction A and 47 state for
reduction B. The fitted parameters are
given in Table 1.  We list the complete calculated
spectrum from both reductions in Table 2.

We turn now to a consideration of how this meson model may be used to treat
in-medium effects on mesons.  In-medium effects may be expected to
modify primarily the long-range part of the interactions.  We therefore focus
our attention on the confining potential in this simple demonstration of
the utility of our meson model.  We use reduction B and the parameters from
the global fit of Table 1 for these examples.

In order to make our calculations independent of a particular model for
the medium effects on the string tension, and to avoid the necessity of
invoking an equilibrium picture of the medium, we plot the meson
masses against $\sigma/\sigma_0$, where $\sigma_0$ is the vacuum value of
the string tension.  Given a model for the behavior of string tension
with respect to a particular variable such as the temperature $T$, this plot
of mass vs. $\sigma/\sigma_0$ can be converted into a plot of mass vs.
temperature.

It is especially interesting to predict whether meson masses fall below
thresholds resulting in the closing of decay channels and
leading to significant modifications in meson
widths.  For this purpose we evaluate the masses
of $\psi''$ and the $D$ meson as a function of decreasing
string tension and display the results in Figure 1.

The calculated $\psi''$ and $D$ masses each differ from the
measured masses by about 30 MeV.  In order to make the curves of Figure 1
pass exactly through the $\psi''$ and $D$ ($\times 2$) measured masses for
$\sigma/\sigma_0=1$, we plot $M(\sigma)M_{exp}/M(\sigma_0)$ on the vertical
axis, where $M(\sigma)$ is the meson mass calculated using the in-medium
string tension $\sigma$, and $M(\sigma_0)$ is the meson mass calculated
using the vacuum value of the string tension $\sigma_0$.

Figure 1 shows that the $\psi''$ mass crosses the $D\bar D$ threshold at a
mass of about 3.71 GeV when the string tenion falls to about $0.9 \sigma_0$.
For string tensions less than this, the avenue for $\psi(3770)$ decay to
$D\bar D$ is blocked.  Since this is by far the dominant decay channel
for the $\psi''$ in the vacuum [8], the model predicts significant
modifications in its decay signatures
will result when medium effects decrease the effective string tension by
more than about 10\%. A similar calculation using reduction A also shows
this ``cross over'' at $\sigma=0.9\sigma_0$.

In Figure 2 we show similar calculations for the masses
of the $\phi$ and the $K\bar K$ threshold as a function of
$\sigma/\sigma_0$.  The figure shows that the $\phi$ mass crosses the
$K\bar K$ threshold at a mass of about 9.8 GeV when $\sigma/\sigma_0
\simeq 0.67$.

The results displayed in Figs. 1 and 2 may be interpreted within the
framework of particular models of medium effects on the string tension.
For example, several methods have been proposed for modeling the
temperature
dependence
of the string tension [9-12] and they exhibit the common feature
of predicting a monotonically decreasing string tension for
increasing temperature.

According to a model for $\sigma(T)$ taken from Ref. [12], as a
particular example,
$$\sigma(T)/\sigma(0)=\sqrt{1-(T/T_c)^2},\eqno(8)$$
where $T_c$ is the critical temperature and $T$ is the temperature of
the
medium.
Applying this model to the $\psi''-D\bar D$ plot in Figure 1, our
calculation predicts that the $\psi''$ will cross the $D\bar D$ threshold
at a temperature of $0.43\ T_c$.

We may compare our result of $0.43\ T_c$ with the value of $0.56\ T_c$
obtained in the study of
Ref. [14] which also used Eq. (8) to model $\sigma(T)$.  This difference may
be attributed to the use of
different constituent quark models.  Ref. [14] used a non-relativistic
Schr\"odinger
equation treatment to compute the $\psi''$ mass, and a bag model to
determine the location of the $D\bar D$ threshold by using
scaling arguments to relate bag pressure to string tension.  Our model
treats all states within a common relativistic framework.

Turning now to the $\phi-K\bar K$ system of Fig. 2, if we again use Eq.
(8)
to model
the temperature dependence of the string tension, we observe that
the $K\bar K$ decay channel of the $\phi$ will close at a temperature of about
$0.74\ T_c$.

We may relate this result to that of a previous
study [15] which used a Nambu-Jona-Lasinio model of in-medium $\phi$ and
$K$
masses.
In this previous study, the $K\bar K$
decay channel for the $\phi$ was found to close at a temperature of
$$T\simeq 0.45\ m_l,\eqno (9)$$
where $m_l$ is
the value of the light constituent quark mass in a free hadron.  Ref. [15]
presents
sample results for light-quark masses in the range 300 MeV $< m_l <$ 600 MeV.
Using the light constituent quark mass obtained in our spectrum fits,
$m_l\simeq 340$ MeV (see Table 1), Eq. (9) predicts that the $K\bar K$
decay channel will become closed for $\phi$ decay at a temperature
of about $150$ MeV. We see that our result of 0.74\ $T_c$ would be comparable
provided we assumed $T_c\simeq 200$ MeV.

Our relativistic in-medium calculations extend
the work of previous non-relativistic potential models by providing a
consistent relativistic framework for evaluating in-medium effects,
through the dependence of our model parameters on properties of the
medium. Work in-progress includes calculations of a large set of vacuum
and in-medium observables for all meson states.

This work was supported in part by the U.S. Department of Energy under
Grant No. DE-FG02-87ER40371, Division of High Energy and Nuclear Physics.
JPV acknowledges support from the Alexander von Humboldt Foundation.  We
acknowledge valuable discussions with J. W. Qiu, S. P. Klevansky, R.
Vogt, J. Hill, F. Wohn, and W. Weise.

\vfill \break

\pagebreak
\raggedright{\bf Table 1.}\\
Best values for the fitted parameters for reductions A, B. Best values for
the seven parameters were determined by a fit to 45 (47) meson states for
reduction A (B).  ``RMS'' is the root-mean-squared
deviation of the fitted meson masses from measured meson masses.  Spectra
are given in Table 2.

\bigskip
\begin{tabular}{l|c|c}
   &A   &  B  \\
\hline
$m_b$ (GeV) &4.65&4.68\\
$m_c$ (GeV) &1.37&1.39\\
$m_s$ (GeV) &0.397&0.405\\
$m_u$ (GeV) &0.339&0.346\\
$\sigma$ (GeV$^2$) &0.233&0.211\\
$\xi$  &0.616&0.444\\
$\rho$ (GeV)  &0.198&0.187\\
RMS (MeV)&43&50\\
\hline\hline
\end{tabular}\\
\pagebreak

\raggedright{\bf Table 2.}\\
Fitted spectra for reductions A and B.  The seven parameters used to obtain
these spectra are given in Table 1.

\pagebreak

\begin{table}\centering
\raggedright{\bf Table 2.}
\begin{tabular}{l|r|r|r}
\hline
Meson     &   M$_{exp}$\ \ &A\ \ &B\ \ \\
\hline
$\pi$&140&-&135\\
$\pi(1300)$&1300&1328&1439\\
 $\pi_2$&1670&1536&1515\\
 $a_1$&1260&1266&1223\\
 $b_1$&1232&1262&1219\\
$\rho$&768&757&812\\
 $a_2$&1318& 1402&1367\\
 $\phi$&1019 &1020&1020\\
 $\phi'$&1680& 1678&1645\\
$f'_2$&1525& 1558&1526\\
$\eta_c(1S)$     &2979  &2975&2993\\
$\eta_c(2S)$     &3590  &3624&3640\\
$\chi_{c0}(1P)$  &3415 &3402&3383\\
$\chi_{c1}(1P)$  &3511 &3486&3461\\
$h_c(1P)$        &3526 &3493&3471 \\
$J/\psi(1S)$     &3097 &3113&3091 \\
$\psi(2S)$       &3686 &3688&3688 \\
$\psi(3770)$     &3770 &3760&3741 \\
$\psi(4040)$     &4040 & 4077&4104 \\
$\psi(4160)$     &4159 &4122&4136 \\
$\psi(4415)$     &4415 & 4415&4456\\
$\chi_{c2}(1P)$  &3556 & 3581&3556 \\
$\chi_{b0}(1P)$  &9860 & 9842&9843 \\
$\chi_{b0}(2P)$  &10232 & 10200&10198\\
$\chi_{b1}(1P)$  &9892 &9860&9863\\
$\chi_{b1}(2P)$  &10255 &10216&10214 \\
$\Upsilon(1S)$   &9460 &9514&9520 \\
$\Upsilon(2S)$   &10023 &9996&9996 \\
$\Upsilon(3S)$   &10355 &10334&10331 \\
$\Upsilon(4S)$   &10580 &10614&10611 \\
$\Upsilon(10860)$&10865 &10861&10860  \\
$\Upsilon(11020)$&11019 &11083&11086 \\
$\chi_{b2}(1P)$  &9913 & 9925&9928\\
$\chi_{b2}(2P)$  &10268 &10271&10270  \\
$K$&494&-&495\\
$K_1$(1406)&1406&1360&1330 \\
$K_1$(1270)&1270 &1336&1287 \\
 $K_2$&1770&1631 &1633\\
 $K^*$&892&902&916 \\
 $K^*_2$&1426&1481&1442 \\
 $D$&1865&1852&1897 \\
 $D^*$(2007)&2007&2034&2004 \\
 $D^*$(2420)&2420&2397&2358 \\
 $D_S$&1971&1928&1968 \\
 $D^*_S$&2110&2108&2076 \\
 $B$&5271&5322&5342 \\
 $B^*$&5352&5328&5347 \\
\hline\hline
\end{tabular}
\end{table}
\pagebreak

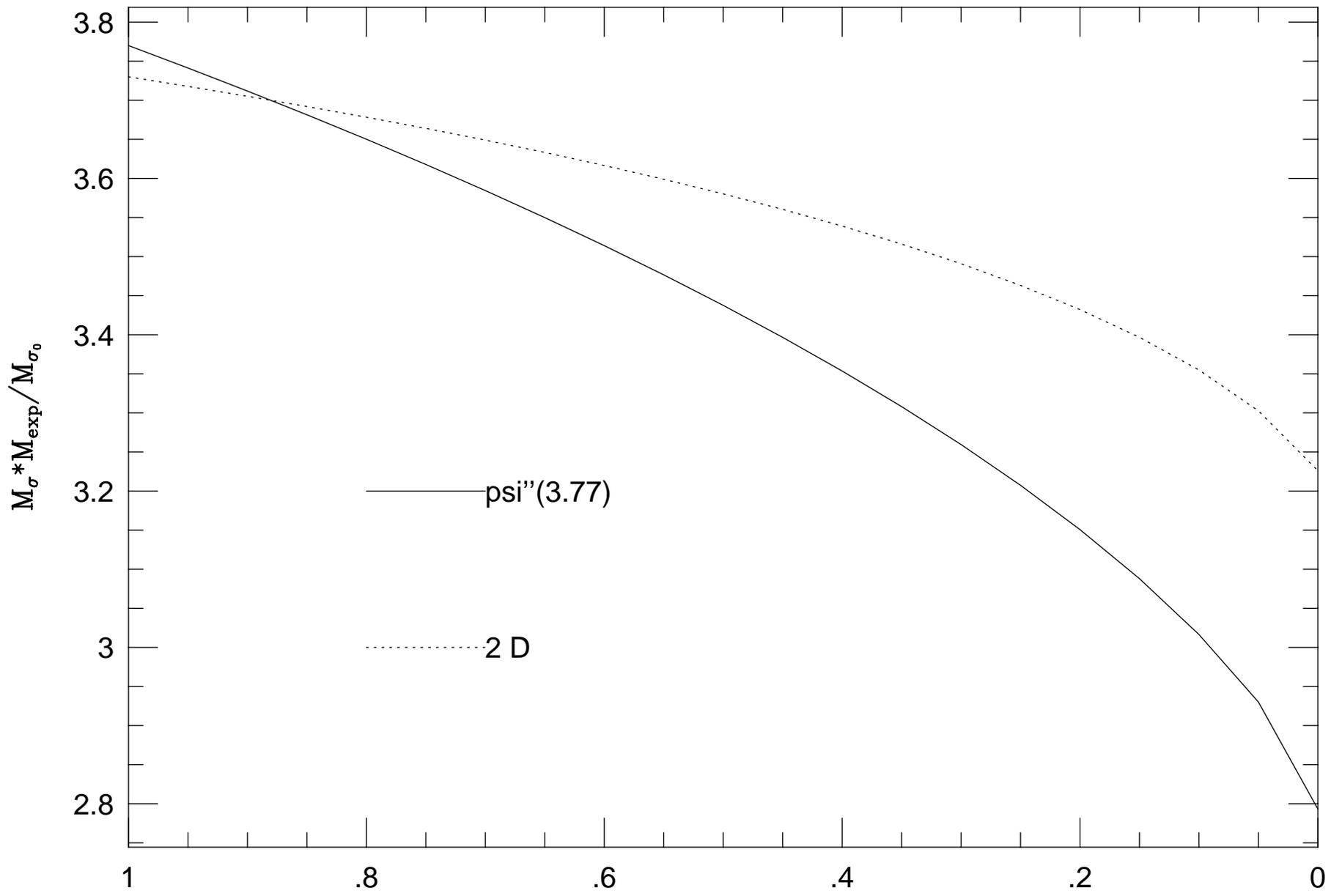
{\bf Figure 1.}\\
Plot of the $\psi''$ mass and the $D\bar D$ threshold as a function of
$\sigma/\sigma_0$, where $\sigma$ is the in-medium value for the string
tension and $\sigma_0$ is the vacuum value of the string tension.  Masses
were calculated with reduction B using the parameters of Table 1.
Since using these
parameters does not yield precisely the experimental vacuum values of the
$\psi''$ and $D$ masses, the vertical axis plots $M(\sigma)M_{exp}/
M(\sigma_0)$ so that the curves pass exactly through
the experimental masses at $\sigma/\sigma_0=1$.

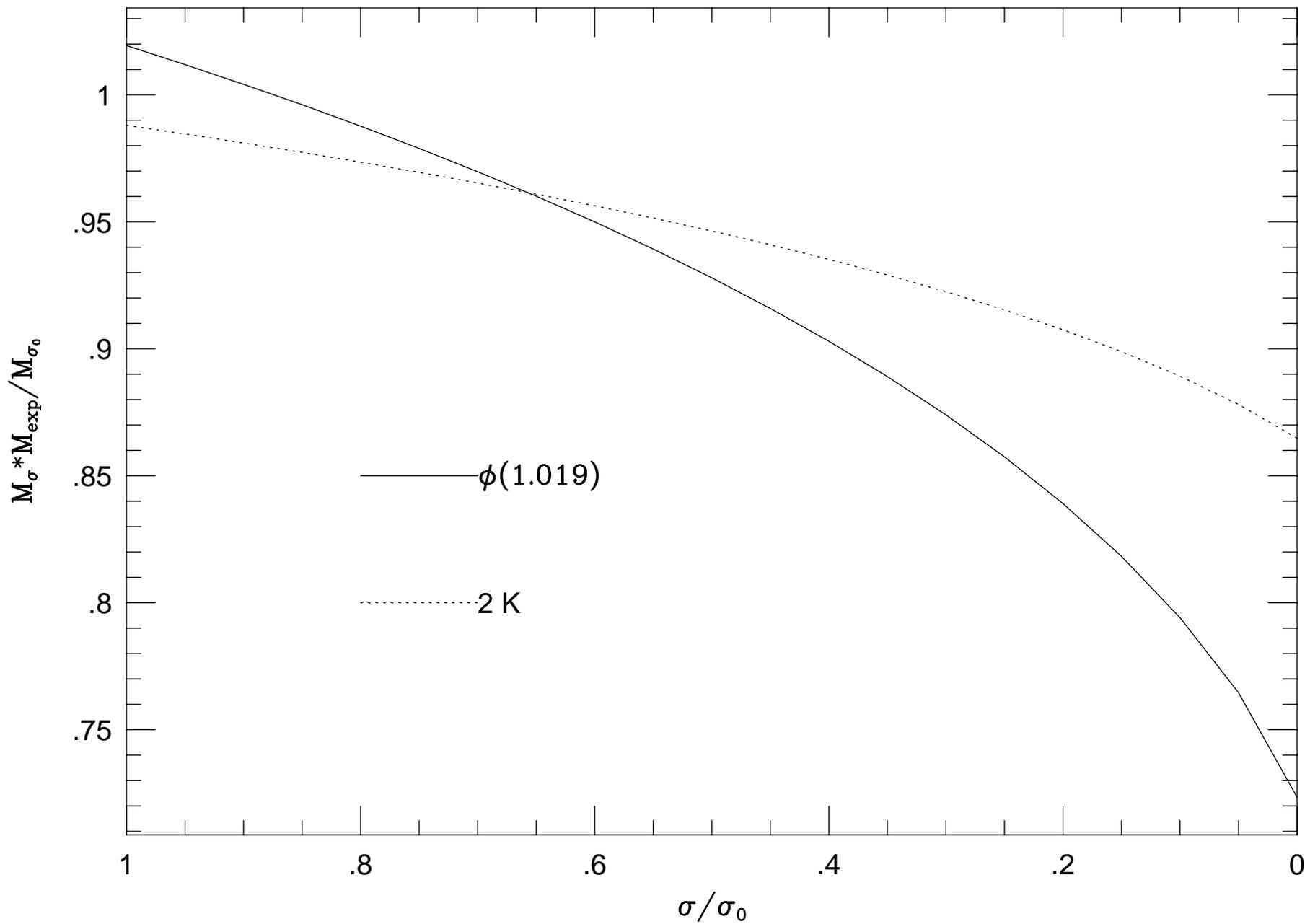
{\bf Figure 2.}\\
Plot of the $\phi$ mass and the $K\bar K$ threshold as a function of
$\sigma/\sigma_0$, where $\sigma$ is the in-medium value for the string
tension and $\sigma_0$ is the vacuum value of the string tension.  Masses
were calculated with reduction B using the parameters of Table 1.
Since using these
parameters does not yield precisely the experimental vacuum values of the
$\phi$ and $K$ masses, the vertical axis plots
$M(\sigma)M_{exp}/M(\sigma_0)$ so that the curves pass exactly through
the experimental values at $\sigma/\sigma_0=1$.

\end{document}